\documentclass[pra,aps,10pt,twocolumn,showpacs,amsmath,amssymb]{revtex4-1}

\usepackage[english]{babel}
\usepackage{afterpage}
\usepackage{graphicx}
\usepackage{dcolumn}
\usepackage{bm}
\usepackage{color}
\usepackage{slashed}
\usepackage{latexsym}

\newcommand{\dd}{\mathrm{d}}         
\newcommand{\ii}{\mathrm{i}}  
\newcommand{\ee}{\mathrm{e}}  

\begin{document}

\title{Radiation pressure in SFA theory: retardation and recoil corrections}
\author{K. Krajewska$^{1,2}$}
\email[E-mail address:\;]{Katarzyna.Krajewska@fuw.edu.pl}
\author{J. Z. Kami\'nski$^1$}
\affiliation{$^1$Institute of Theoretical Physics, Faculty of Physics, University of Warsaw, Pasteura 5,
02-093 Warszawa, Poland \\
$^2$Department of Physics and Astronomy, University of Nebraska, Lincoln, Nebraska 68588-0299, USA
}

\date{\today}

\begin{abstract}
Radiation pressure effects in ionization by short linearly-polarized laser pulses are investigated in the framework of strong-field approximation, in 
both nonrelativistic and relativistic formulations. Differences between both approaches are discussed, and retardation and recoil corrections are defined. 
It is demonstrated how these corrections can be incorporated into the nonrelativistic approach, leading to the so-called quasi-relativistic formulation. These three 
approaches are further applied to the analysis of signatures of radiation pressure in energy-angular distributions of photoelectrons. It is demonstrated that, for Ti:Sapphire laser 
pulses of intensities up to $10^{16}\mathrm{W/cm}^2$, predictions of the quasi-relativistic formulation agree well with those of the full relativistic one, 
and that the recoil corrections contribute predominantly to the radiation pressure effects.
\end{abstract}


\maketitle

\section{Introduction}
\label{sec::intro}

Effects in strong-field ionization of atoms attributed to the radiation pressure~\cite{press0} have been studied recently both experimentally~\cite{press1} and theoretically~\cite{press2,press3,Chelkowski,press4}. 
It is well-known that these effects are related to the photon momentum being transferred to the atomic system. Hence, modifying the distributions of ionized electrons. 
Such a phenomenon is absent in the nonrelativistic theoretical approach because of the dipole approximation applied to the laser field. 

Except for the Ivanov's paper~\cite{press4}, where the numerical solution of the Dirac equation coupled to the 2-cycle laser pulse has been considered, in Refs.~\cite{press2,Chelkowski,press3} 
different types of the Strong-Field Approximation (SFA) have been applied that go beyond the dipole approximation. Not surprisingly, in all cases the radiation pressure effects have become visible. 
The aim of this paper is to further develop the SFA approach and to analyze angular distributions of photoelectrons ionized by linearly-polarized few-cycle pulses. This is in contrast to previous
works~\cite{press1,press2,press3,Chelkowski,press4} where circularly-polarized fields were considered in order to describe multiphoton ionization. Our numerical analysis concerns the multiphoton
ionization of hydrogen atoms, even though we expect that similar effects should be observed for heavier atomic systems.

This paper is organized as follows. In Sec.~\ref{sec:laser}, we will define properties of linearly-polarized and short laser pulses. For such pulses, in Sec.~\ref{sec:theory} we will formulate 
three types of SFA approaches to ionization, each in the velocity gauge. In Secs.~\ref{sec:angle1},~\ref{sec:angle2}, and~\ref{sec:angle3}, we shall compare predictions of these 
approaches and investigate effects related to the radiation pressure in angular distributions of photoelectrons. In Sec.~\ref{sec:Conclusions}, we will draw some conclusions and formulate prospects for further investigations.

In this paper we keep $\hbar=1$. Hence, the fine-structure constant equals $\alpha=e^2/(4\pi\varepsilon_0c)$. In our numerical analysis we use relativistic units (rel. units) such that 
$\hbar=m_{\rm e}=c=1$ where $m_{\rm e}$ is the electron rest mass. Furthermore, we denote the product of any two four-vectors $a^{\mu}$ and $b^{\mu}$ as $a\cdot b = a^{\mu}b_{\mu}=a^0b^0-a^1b^1-a^2b^2-a^3b^3$ 
($\mu = 0,1,2,3$), where the Einstein summation convention is used. We employ the Feynman notation $\slashed{a} = \gamma\cdot a=\gamma^{\mu} a_{\mu}$ for the contraction with the Dirac matrices 
$\gamma^{\mu}$ and we use a customary notation $\bar{u}=u^{\dagger}\gamma^0$, where $u^{\dagger}$ is the Hermitian conjugate of a bispinor $u$. For the four-vectors we use both the contravariant 
$(a^0,a^1,a^2,a^3)$ and the standard $(a_0,a_x,a_y,a_z)=(a_0,\bm{a})$ notations. Finally, we use the light-cone variables. 
For a given space direction determined by a unit vector $\bm{n}$ (which in this paper denotes the direction of the pulse propagation) and for an arbitrary four-vector $a$, we define these variables
such that $a^{\|}=\bm{n}\cdot\bm{a}$, $a^-=a^0-a^{\|}$, $a^+=(a^0+a^{\|})/2$, and $\bm{a}^{\bot}=\bm{a}-a^{\|}\bm{n}$. Thus, the relativistic scalar product is 
$a\cdot b=a^+b^-+a^-b^+-\bm{a}^{\bot}\cdot\bm{b}^{\bot}$ and $\dd^4x=\dd x^+\dd x^-\dd^2x^{\bot}$.

\section{Laser pulse description}
\label{sec:laser}

While in prior investigations of radiation pressure effects in multiphoton ionization~\cite{press1,press2,press3,Chelkowski,press4} the circularly-polarized laser fields have been considered,
in this work we consider the laser field of a linear polarization. We assume that the laser pulse lasts for time $T_{\mathrm{p}}$, which defines its fundamental frequency 
$\omega=2\pi/T_{\mathrm{p}}$ (being the smallest frequency in the Fourier decomposition of the pulse). We also assume that the plane-wave-fronted pulse approximation can be used, 
which means that the electromagnetic field is a function of the phase $\phi=k\cdot x=\omega t-\bm{k}\cdot\bm{x}$, $0<\phi<2\pi$. We choose the system of coordinates such that the laser pulse 
propagates in the $z$-direction, $\bm{k}=(\omega/c)\bm{n}=(\omega/c)\bm{e}_z$, and its linear polarization vector is along the $x$-axis. This means that the electric field has the form
\begin{equation}
\bm{\mathcal{E}}(\phi)=\mathcal{E}_0f_{\mathcal{E}}(\phi)\bm{e}_x,
\label{las1}
\end{equation}
where $\mathcal{E}_0>0$. The so-called shape function for the electromagnetic field, $f_{\mathcal{E}}(\phi)$, vanishes for $\phi <0$ and $\phi > 2\pi$, and have a continuous first derivative 
for all real $\phi$. In our numerical analysis, we use the function with the sine-squared envelope,
\begin{equation}
f_{\mathcal{E}}(\phi)=N_0\sin^2\Bigl(\frac{\phi}{2}\Bigr)\sin(N_{\mathrm{osc}}\phi),
\label{las2}
\end{equation}
where $N_{\mathrm{osc}}$ defines the number of cycles in the pulse, and the positive real constant $N_0$ is adjusted such that the normalization condition:
\begin{equation}
\frac{1}{2\pi}\int_0^{2\pi}\dd\phi [f_{\mathcal{E}}(\phi)]^2=\frac{1}{2}N^2_{\mathrm{osc}}
\label{las3}
\end{equation}
holds. With this normalization, the time-averaged intensity of the laser pulse equals
\begin{equation}
I=I_{\mathrm{rel}}\Bigl(\frac{\omega_{\mathrm{L}}}{m_{\mathrm{e}}c^2}\Bigr)^2\mu^2,
\label{las4}
\end{equation}
where $\omega_{\mathrm{L}}=N_{\mathrm{osc}}\omega$ is the laser carrier frequency, $I_{\mathrm{rel}}$ is the relativistic unit of intensity,
\begin{equation}
I_{\mathrm{rel}}=\frac{m_{\mathrm{e}}^4c^6}{8\pi\alpha}\approx 2.3\times 10^{29}\,\mathrm{W/cm}^2,
\label{las5}
\end{equation}
and the dimensionless and relativistically invariant parameter,
\begin{equation}
\mu=\frac{|e\mathcal{E}_0|}{m_{\mathrm{e}}c\omega}=N_{\mathrm{osc}}\frac{|e\mathcal{E}_0|}{m_{\mathrm{e}}c\omega_{\mathrm{L}}},
\label{las6}
\end{equation}
called sometimes the Ritus parameter~\cite{ritus1,ritus2} (see, also the review articles~\cite{FKK,PiazzaRev}), marks the transition between the nonrelativistic 
($\mu \ll 1$) and ultra-relativistic ($\mu\gg 1$) regimes of laser-matter interactions. Next, we introduce the function
\begin{equation}
f(\phi)=-\int_0^{\phi}\dd\phi'\, f_{\mathcal{E}}(\phi'),
\label{las7}
\end{equation}
which also vanishes for $\phi<0$ and $\phi>2\pi$, and it defines the electromagnetic vector potential,
\begin{equation}
\bm{A}(\phi)=A_0f(\phi)\bm{e}_x=\frac{\mathcal{E}_0}{\omega}f(\phi)\bm{e}_x.
\label{las8}
\end{equation}
In the following, we will use the relativistic notation such that $A(\phi)=A_0f(\phi)\varepsilon$, where $\varepsilon=(0,\bm{e}_x)$ and $\varepsilon\cdot\varepsilon=-1$. 
In the nonrelativistic case when the dipole approximation is used, meaning that the vector potential and the electric field depend only on time, we will assume $\phi=k^0x^0=\omega t$.

\begin{figure}
\includegraphics[width=7cm]{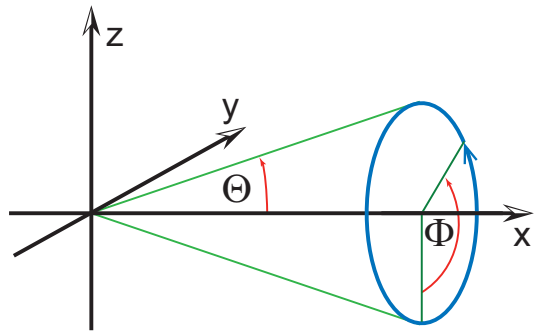}
\caption{(Color online) Polar $\Theta$ and azimuthal $\Phi$ angles in the Cartesian coordinate system, defined by the triad of unit vectors $(-\bm{e}_z,\bm{e}_y,\bm{e}_x)$. 
}
\label{pedcone}
\end{figure}

In our analysis, we shall use the right-hand-side system of Cartesian coordinates defined by the triad of orthonormal vectors $(-\bm{e}_z,\bm{e}_y,\bm{e}_x)$. 
In this system, we introduce the polar $\Theta$ and azimuthal $\Phi$ angles (cf. Fig.~\ref{pedcone}). Thus, an arbitrary vector $\bm{p}$ is defined by its length $|\bm{p}|$ 
and spherical angles $(\Theta_{\bm{p}},\Phi_{\bm{p}})$ such that
\begin{equation}
\bm{p}=|\bm{p}|[(-\bm{e}_z\cos\Phi_{\bm{p}}+\bm{e}_y\sin\Phi_{\bm{p}})\sin\Theta_{\bm{p}}+\bm{e}_x\cos\Theta_{\bm{p}}]
\label{las9}
\end{equation}
and
\begin{equation}
\dd^3p=\bm{p}^2\dd|\bm{p}|\dd^2\Omega_{\bm{p}}, \quad \dd^2\Omega_{\bm{p}}=\sin\Theta_{\bm{p}}\dd\Theta_{\bm{p}}\dd\Phi_{\bm{p}}.
\label{las10}
\end{equation}
This choice of coordinates is motivated by the symmetry of the ionization probability distribution, which in the nonrelativistic limit and for the linearly-polarized laser field in the $x$-direction 
is independent of the azimuthal angle $\Phi_{\bm{p}}$, where $\bm{p}$ is the momentum of photoelectrons. This suggests that any deviations from this symmetry can be attributed to the radiation pressure.

\section{SFA theories and approximations}
\label{sec:theory}

The SFA in strong-field ionization has been introduced in the seminal Keldysh paper~\cite{Keldysh}, and during the last 50 years it has been formulated in different ways and for various purposes (see, 
e.g., Refs.~\cite{Perelomov,Gribakin,Karnakov,Popruzh,CKK,Faisal,Reiss,Sujata1}). In the present investigations we choose the SFA in the velocity gauge, \textit{only} because in the prior theoretical studies of the radiation pressure effects \cite{press2,press3,Chelkowski,press4} exactly this gauge (up to the Kramers-Henneberger transformation) has been considered. An extended discussion of these effects in the length gauge is going to be presented in due course.

In the following Sections, we will formulate three versions of SFA in the velocity gauge. In Sec.~\ref{sec:NSFA} the Nonrelativistic SFA (NSFA) will be presented, which then is followed by the 
fully Relativistic SFA (RSFA) introduced in Sec.~\ref{sec:RSFA}. It is possible to reformulate the NSFA such that it accounts for the interaction of photoelectrons with their parent ions 
(for instance, in terms of the complex-time trajectories, e.g.,~\cite{Karnakov,Popruzh,CKK}). For the RSFA it is not an easy task, although some attempts have been already 
undertaken~\cite{eikHeidelberg,eikHeidelberg2} (see, also the theoretical analysis for the Klein-Gordon equation~\cite{Kam}). For this reason, we will formulate in Sec.~\ref{sec:QRSFA} 
the Quasi-Relativistic SFA (QRSFA) which computationally is not so demanding and, as being very similar to the NSFA, could be further generalized in terms of the 
complex-time trajectories. One of the main purposes of this paper is to compare predictions of the QRSFA and the RSFA, and to determine for which laser pulse intensities the former approach is applicable.

A general scheme for the SFA is as follows. We split the exact time-dependent Hamiltonian,
\begin{equation}
H(t)=H_0+H_I(t),
\label{sfa1}
\end{equation}
into the static, $H_0$, and the time-dependent, $H_I(t)$, parts. The latter accounts for the laser-matter interaction and it vanishes asymptotically (i.e., it vanishes in the remote past and future). 
For $H_0$, we define the bound state $|\psi_0\rangle$ of energy $E_0$. Then, the exact probability amplitude for the transition from the bound state $|\psi_0\rangle$ 
to the scattering state $|\psi_{\bm p}(t)\rangle$, labeled by the asymptotic momentum $\bm{p}$, is equal to
\begin{equation}
\mathcal{A}(\bm{p})=-\ii\int \dd t \langle\psi_{\bm{p}}(t)|H_I(t)|\psi_0\rangle \ee^{-\ii E_0t}.
\label{sfa2}
\end{equation}
$|\psi_{\bm{p}}(t)\rangle$ is the exact solution for the full Hamiltonian $H(t)$ which, for $t\rightarrow\infty$, tends to the specific scattering state with the incoming spherical waves 
defined for the Hamiltonian $H_0$. This state is very difficult to determine, therefore some approximations are applied. The SFA consists in replacing this exact solution by the 
so-called Volkov solution~\cite{Volkov}. In this approximation, the influence of the binding potential is neglected, i.e., the probability amplitude in the SFA becomes,
\begin{equation}
\mathcal{A}_{\mathrm{SFA}}(\bm{p})=-\ii\int \dd t \langle\psi^{(0)}_{\bm{p}}(t)|H_I(t)|\psi_0\rangle \ee^{-\ii E_0t}.
\label{sfa3}
\end{equation}
Here, the Volkov solution $|\psi^{(0)}_{\bm{p}}(t)\rangle$ fulfills the Schr\"odinger equation,
\begin{equation}
\ii \partial_t |\psi^{(0)}_{\bm{p}}(t)\rangle=[H_{\mathrm{kin}}+H_I(t)]|\psi^{(0)}_{\bm{p}}(t)\rangle,
\label{sfa4}
\end{equation}
where $H_{\mathrm{kin}}$ is that part of $H_0$ which remains after neglecting the binding potential. This is equivalent to treating the exact solution $|\psi_{\bm p}(t)\rangle$ 
in the lowest Born approximation with respect to the binding potential. Such an approximation is justified if the kinetic energy of photoelectrons is sufficiently large compared to the binding energy $E_0$, 
or if the action of the laser pulse exceeds the influence of the binding potential for photoelectrons.

In the following discussion, in order to trace similarities between the nonrelativistic and relativistic approaches, we shall use the relativistic notation in both cases. 
For this purpose, we rewrite Eq.~\eqref{sfa3} as follows,
\begin{align}
\mathcal{A}_{\mathrm{SFA}}(\bm{p})=-\ii\int\frac{\dd^3 q}{(2\pi)^3}\int \dd^4 x \bigl[&\psi^{(0)}_{\bm{p}}(x)\bigr]^{\dagger}\frac{1}{c}H_I(x) \nonumber \\
\times &\tilde{\psi}_0(\bm{q}) \ee^{-\ii q\cdot x},
\label{sfa5}
\end{align}
where
\begin{equation}
\psi_0(\bm{x})=\int\frac{\dd^3 q}{(2\pi)^3}\,\ee^{\ii \bm{q}\cdot\bm{x}}\tilde{\psi}_0(\bm{q})
\label{sfa6}
\end{equation}
and $q=(q^0,\bm{q})$ with $q^0=E_0/c$. In the following, we shall call such four-component objects the `four-vectors' or the `four-momenta'. One has to remember, however, that
in general they are not the true relativistic four-vectors as they do not transform properly under the Lorentz boosts.

\subsection{NSFA}
\label{sec:NSFA}

Let us now apply the general SFA scheme to the nonrelativistic ionization described in the velocity gauge. In this case, the laser-electron interaction takes the form
\begin{equation}
\frac{1}{c}H_I(x)=\ii\frac{e}{m_{\mathrm{e}}c}\bm{A}(k^0x^0)\cdot\bm{\nabla}+\frac{e^2}{2m_{\mathrm{e}}c}\bm{A}^2(k^0x^0),
\label{nsfa1}
\end{equation}
whereas the Volkov solution is defined as
\begin{align}
\psi^{(0)}_{\bm{p}}(x)=\frac{1}{\sqrt{V}}\exp\Bigl[-\ii p\cdot x-&\ii\int_0^{k^0x^0}\dd\phi'\Bigl(\frac{eA(\phi')\cdot p}{m_{\mathrm{e}}ck^0}  \nonumber \\
&-\frac{e^2A^2(\phi')}{2m_{\mathrm{e}}ck^0}\Bigr)\Bigr].
\label{nsfa2}
\end{align}
Here, $V$ is the quantization volume and $p=(p^0,\bm{p})$ with $p^0=\bm{p}^2/(2m_{\mathrm{e}}c)$.

For the linearly-polarized laser pulse defined in Sec.~\ref{sec:laser}, the probability amplitude of ionization under the NSFA becomes
\begin{align}
\mathcal{A}_{\mathrm{NSFA}}&(\bm{p})=\ii\frac{\tilde{\psi}_0(\bm{p})}{\sqrt{V}}\int_0^{2\pi}\dd\phi \Bigl[\frac{m_{\mathrm{e}}c\mu}{m_{\mathrm{e}}ck^0}\varepsilon\cdot p f(\phi) \nonumber \\
-&\frac{(m_{\mathrm{e}}c\mu)^2}{2m_{\mathrm{e}}ck^0}f^2(\phi)\Bigr]\exp\Bigl[\ii\frac{\bar{p}^0-q^0}{k^0}\phi+\ii G_{\mathrm{NSFA}}(\phi)\Bigr].
\label{nsfa3}
\end{align}
Note that in order to trace similarities with the full relativistic treatment, in this and in the following formulas we keep the factors $m_{\mathrm{e}}c$.
In Eq.~\eqref{nsfa3}, we have introduced the laser-dressed energy (divided by $c$) of the final electron,
\begin{equation}
\bar{p}^0=p^0+\Bigl[-\frac{m_{\mathrm{e}}c\mu}{m_{\mathrm{e}}ck^0}\varepsilon\cdot p \langle f\rangle+\frac{(m_{\mathrm{e}}c\mu)^2}{2m_{\mathrm{e}}ck^0}\langle f^2\rangle\Bigr]k^0,
\label{nsfa4}
\end{equation}
along with
\begin{align}
G_{\mathrm{NSFA}}(\phi)=\int_0^{\phi}\dd\phi' & \Bigl[\frac{m_{\mathrm{e}}c\mu}{m_{\mathrm{e}}ck^0}\varepsilon\cdot p \bigl(f(\phi)-\langle f\rangle\bigr)
\nonumber \\
&+\frac{(m_{\mathrm{e}}c\mu)^2}{2m_{\mathrm{e}}ck^0}\bigl(f^2(\phi)-\langle f^2\rangle\bigr)\Bigr].
\label{nsfa5}
\end{align}
Here,
\begin{equation}
\langle f\rangle=\frac{1}{2\pi}\int_0^{2\pi}\dd\phi f(\phi)
\label{nsfa6}
\end{equation}
and, similarly, for $\langle f^2\rangle$. Introducing now the Fourier expansions~\cite{KKC,KKBW,KK1},
\begin{equation}
\bigl[f(\phi)\bigr]^j\exp\bigl[\ii G_{\mathrm{NSFA}}(\phi)\bigr]=\sum_{N=-\infty}^{\infty}G_{\mathrm{NSFA},N}^{(j)}\ee^{-\ii N\phi},
\label{nsfa7}
\end{equation}
for $j=1,2$, we arrive at the expression for the probability amplitude of ionization such that
\begin{equation}
\mathcal{A}_{\mathrm{NSFA}}(\bm{p})=\ii\frac{m_{\mathrm{e}}c\mu}{\sqrt{V}}\mathcal{D}_{\mathrm{NSFA}}(\bm{p}),
\label{nsfa8}
\end{equation}
where
\begin{align}
\mathcal{D}_{\mathrm{NSFA}}(\bm{p})=&\sum_{N=-\infty}^{\infty}\frac{\ee^{2\pi \ii (\bar{p}^0-q^0-Nk^0)/k^0}-1}{\ii (\bar{p}^0-q^0-Nk^0)/k^0}
\nonumber \\
\times&\Bigl[\frac{\varepsilon\cdot p}{m_{\mathrm{e}}ck^0}G_{\mathrm{NSFA},N}^{(1)}+\frac{m_{\mathrm{e}}c\mu}{2m_{\mathrm{e}}ck^0}G_{\mathrm{NSFA},N}^{(2)}\Bigr]\tilde{\psi}_0(\bm{p}).
\label{nsfa9}
\end{align}
Based on these expressions, we define the energy-angular differential probability distribution of ionization,
\begin{equation}
\frac{\dd^3 P_{\mathrm{NSFA}}}{\dd E_{\mathrm{p}}\dd^2\Omega_{\bm{p}}}=\Bigl(\frac{m_{\mathrm{e}}c\mu}{2\pi}\Bigr)^2\frac{m_{\mathrm{e}}|\bm{p}|}{2\pi}|\mathcal{D}_{\mathrm{NSFA}}(\bm{p})|^2,
\label{nsfa10}
\end{equation}
and its dimensionless equivalent
\begin{equation}
\mathcal{P}_{\mathrm{NSFA}}(\bm{p})=\mathcal{P}_{\mathrm{NSFA}}(E_{\bm{p}},\Theta_{\bm{p}},\Phi_{\bm{p}})=\alpha^2m_{\mathrm{e}}c^2\frac{\dd^3 P_{\mathrm{NSFA}}}{\dd E_{\mathrm{p}}\dd^2\Omega_{\bm{p}}},
\label{nsfa11}
\end{equation}
or
\begin{equation}
\mathcal{P}_{\mathrm{NSFA}}(\bm{p})=\alpha^2\frac{(m_{\mathrm{e}}c)^4\mu^2}{(2\pi)^3}|\bm{p}|\cdot|\mathcal{D}_{\mathrm{NSFA}}(\bm{p})|^2,
\label{nsfa12}
\end{equation}
which is the probability distribution in the atomic units.
The latter being presented in our numerical illustrations.

In the following Sections, we will derive other versions of the SFA and, if it does not create misunderstandings, we will omit the subscripts labeling these formulations.

\subsection{RSFA}
\label{sec:RSFA}

In the relativistic theory the procedure is very similar. The laser-electron interaction Hamiltonian in the velocity gauge has the form,
\begin{equation}
H_I(x)=ec\gamma^0\slashed{A}(k\cdot x),
\label{rsfa1}
\end{equation}
and the corresponding Volkov solution (note that the superscript $(+)$ denotes solutions of the Dirac equation for electrons, i.e., for particles of positive energies) becomes (see, e.g., \cite{FKK,PiazzaRev}),
\begin{equation}
\psi^{(+)}_{\bm{p}\lambda}(x)=\sqrt{\frac{m_{\mathrm{e}}c^2}{VE_{\bm{p}}}}\Bigl(1+\frac{m_{\mathrm{e}}c\mu}{2p\cdot k}f(k\cdot x)\slashed{\varepsilon}\slashed{k}\Bigr)\ee^{-\ii S_{\bm{p}}(x)}u^{(+)}_{\bm{p}\lambda},
\label{rsfa2}
\end{equation}
where
\begin{equation}
S_{\bm{p}}(x)=p\cdot x-\int_0^{k\cdot x}\dd\phi\Bigl[\frac{m_{\mathrm{e}}c\mu}{p\cdot k} f(\phi)\varepsilon\cdot p-\frac{(m_{\mathrm{e}}c\mu)^2}{2p\cdot k} f^2(\phi)\Bigr]
\label{rsfa3}
\end{equation}
and
\begin{equation}
p=(E_{\bm{p}}/c,\bm{p}), \quad E_{\bm{p}}=\sqrt{(c\bm{p})^2+(m_{\mathrm{e}}c^2)^2}.
\label{rsfa4}
\end{equation}
The free-electron Dirac bispinor $u^{(+)}_{\bm{p}\lambda}$ fulfills the system of linear algebraic equations,
\begin{equation}
(\slashed{p}-m_{\mathrm{e}}c)u^{(+)}_{\bm{p}\lambda}=0,
\label{rsfa5}
\end{equation}
and satisfies the normalization condition $\bar{u}^{(+)}_{\bm{p}\lambda}u^{(+)}_{\bm{p}\lambda'}=\delta_{\lambda\lambda'}$, with $\lambda=\pm 1$ labeling the spin degrees of freedom~\cite{BjorkenDrell}.

For our further purposes we define the $B$-functions,
\begin{align}
B^{(0)}_{\bm{p}\lambda,\lambda_{\mathrm{i}}}(\bm{q})=&\bar{u}^{(+)}_{\bm{p}\lambda}\slashed{n}\tilde{\psi}_0(\bm{q},\lambda_{\mathrm{i}}),\label{rsfa6a} \\
\frac{\varepsilon\cdot p}{p\cdot n}B^{(1)}_{\bm{p}\lambda,\lambda_{\mathrm{i}}}(\bm{q})=&\bar{u}^{(+)}_{\bm{p}\lambda}\slashed{\varepsilon}\tilde{\psi}_0(\bm{q},\lambda_{\mathrm{i}}),
\label{rsfa6}
\end{align}
where $\psi_0(\bm{x},\lambda_{\mathrm{i}})$ is the bound state solution of the Dirac equation of energy $cq^0=E_0<m_{\mathrm{e}}c^2$, with the spin polarization $\lambda_{\mathrm{i}}$. The explicit forms of these solutions for the ground state of the hydrogen-like atoms are presented in \cite{BjorkenDrell}. Since $\slashed{\varepsilon}\slashed{k}\slashed{\varepsilon}=\slashed{k}$, we arrive at the following expression for the relativistic probability amplitude,
\begin{align}
\mathcal{A}_{\mathrm{RSFA}}(\bm{p},\lambda;\lambda_{\mathrm{i}})&=\ii\sqrt{\frac{m_{\mathrm{e}}c^2}{VE_{\bm{p}}}}\int\frac{\dd^3q}{(2\pi)^3}\int\dd^4x\,\ee^{\ii S_{\bm{p}}(x)-\ii q\cdot x}
\nonumber \\
&\times\Bigl[\frac{m_{\mathrm{e}}c\mu}{p\cdot n}B^{(1)}_{\bm{p}\lambda,\lambda_{\mathrm{i}}}(\bm{q}) f(k\cdot x) \varepsilon\cdot p
\nonumber \\
&\quad -\frac{(m_{\mathrm{e}}c\mu)^2}{2p\cdot n}B^{(0)}_{\bm{p}\lambda,\lambda_{\mathrm{i}}}(\bm{q})f^2(k\cdot x) \Bigr].
\label{rsfa7}
\end{align}
Now, by analogy to Eq.~\eqref{nsfa4}, we introduce the laser-dressed electron final momentum,
\begin{equation}
\bar{p}=p+\Bigl[-\frac{m_{\mathrm{e}}c\mu}{p\cdot k}\langle f\rangle\varepsilon\cdot p +\frac{(m_{\mathrm{e}}c\mu)^2}{2p\cdot k}\langle f^2\rangle\Bigr]k,
\label{rsfa8}
\end{equation}
which fulfills the conditions (expressed in the light-cone coordinates): $\bar{p}^-=p^-$ and $\bar{\bm{p}}^{\bot}=\bm{p}^{\bot}$. This allows us to write,
\begin{align}
S_{\bm{p}}(x)- q\cdot x=&(p^--q^-)x^+-(\bm{p}^{\bot}-\bm{q}^{\bot})\cdot\bm{x}^{\bot} 
\nonumber \\
&+(\bar{p}^+-q^+)x^-+G_{\mathrm{RSFA}}(k^0x^-),
\label{rsfa9}
\end{align}
where
\begin{align}
G_{\mathrm{RSFA}}(\phi)=\int_0^{\phi}\dd\phi'\Bigl[&-\frac{m_{\mathrm{e}}c\mu}{p\cdot k}\bigl(f(\phi)-\langle f\rangle\bigr)\varepsilon\cdot p
\nonumber \\
&+\frac{(m_{\mathrm{e}}c\mu)^2}{2p\cdot k}\bigl(f^2(\phi)-\langle f^2\rangle\bigr) \Bigr].
\label{rsfa10}
\end{align}
Thus, the integrations over $x^+$ and $\bm{x}^{\bot}$ lead to the conservation conditions $p^-=q^-$ and $\bm{p}^{\bot}=\bm{q}^{\bot}$. Their solution can be put in the form
\begin{equation}
\bm{q}=\bm{Q}=\bm{p}+(q^0-p^0)\bm{n}.
\label{rsfa11}
\end{equation}
After performing the integration $\dd^3 q$ in Eq.~\eqref{rsfa7}, we end up with the expression for the relativistic probability amplitude of ionization,
\begin{align}
\mathcal{A}_{\mathrm{RSFA}}(\bm{p},\lambda;\lambda_{\mathrm{i}})&=\ii\sqrt{\frac{m_{\mathrm{e}}c^2}{VE_{\bm{p}}}}\int_0^{2\pi}\dd\phi\,\ee^{\ii (\bar{p}^0-q^0)x^-+\ii G_{\mathrm{RSFA}}(\phi)}
\nonumber \\
&\times\Bigl[\frac{m_{\mathrm{e}}c\mu}{p\cdot k}B^{(1)}_{\bm{p}\lambda,\lambda_{\mathrm{i}}}(\bm{Q}) f(k\cdot x) \varepsilon\cdot p
\nonumber \\
&\quad -\frac{(m_{\mathrm{e}}c\mu)^2}{2p\cdot k}B^{(0)}_{\bm{p}\lambda,\lambda_{\mathrm{i}}}(\bm{Q})f^2(k\cdot x) \Bigr].
\label{rsfa12}
\end{align}
Following the same steps as above, we introduce the Fourier decomposition~\cite{KKC,KKBW,KK1},
\begin{equation}
\bigl[f(\phi)\bigr]^j\exp\bigl[\ii G_{\mathrm{RSFA}}(\phi)\bigr]=\sum_{N=-\infty}^{\infty}G_{\mathrm{RSFA},N}^{(j)}\ee^{-\ii N\phi},
\label{rsfa13}
\end{equation}
and arrive at
\begin{equation}
\mathcal{A}_{\mathrm{RSFA}}(\bm{p},\lambda;\lambda_{\mathrm{i}})=\ii\frac{m_{\mathrm{e}}c\mu}{\sqrt{V}}\mathcal{D}_{\mathrm{RSFA}}(\bm{p},\lambda;\lambda_{\mathrm{i}}),
\label{rsfa14}
\end{equation}
with
\begin{align}
\mathcal{D}_{\mathrm{RSFA}}(\bm{p},\lambda;\lambda_{\mathrm{i}})=&\sum_{N=-\infty}^{\infty}\frac{\ee^{2\pi \ii (\bar{p}^0-q^0-Nk^0)/k^0}-1}{\ii (\bar{p}^0-q^0-Nk^0)/k^0}
\nonumber \\
\times&\Bigl[\frac{\varepsilon\cdot p}{p\cdot k}G_{\mathrm{RSFA},N}^{(1)}B^{(1)}_{\bm{p}\lambda,\lambda_{\mathrm{i}}}(\bm{Q}) \nonumber \\
& +\frac{m_{\mathrm{e}}c\mu}{2p\cdot k}G_{\mathrm{RSFA},N}^{(2)}B^{(0)}_{\bm{p}\lambda,\lambda_{\mathrm{i}}}(\bm{Q})\Bigr].
\label{rsfa15}
\end{align}
Hence, the dimensionless probability distribution of ionization can be calculated [cf. the comment below Eq.~\eqref{nsfa12}],
\begin{equation}
\mathcal{P}_{\mathrm{RSFA}}(\bm{p})=\alpha^2\frac{(m_{\mathrm{e}}c)^4\mu^2}{(2\pi)^3}|\bm{p}|\Bigl[\frac{1}{2}\sum_{\lambda,\lambda_{\mathrm{i}}=\pm}|\mathcal{D}_{\mathrm{RSFA}}(\bm{p},\lambda;\lambda_{\mathrm{i}})|^2\Bigr],
\label{rsfa16}
\end{equation}
where the summation over the final, $\lambda$, and averaging over the initial, $\lambda_{\mathrm{i}}$, electron spin degrees of freedom have been carried out. 
This formula is the one that should be compared with its nonrelativistic version~\eqref{nsfa12}.

\subsection{QRSFA}
\label{sec:QRSFA}

Based on the full relativistic treatment, it is now possible to modify the nonrelativistic theory so it accounts for the radiation pressure effects. In doing so, we shall follow the proposal put forward 
by Ehlotzky~\cite{Ehlo} (see, also Ref.~\cite{Chelkowski} and references therein), and discussed in the context of the laser-assisted scattering processes~\cite{EJK}. We call this approach the quasi-relativistic SFA, as it only accounts for linear terms 
in the nonrelativistic expansion with respect to $1/c$. For the QRSFA, the modifications of the NSFA are done in two places: in the laser-matter interaction Hamiltonian,
\begin{equation}
\frac{1}{c}H_I(x)=\ii\frac{e\bm{A}(k\cdot x)\cdot\bm{\nabla}}{m_{\mathrm{e}}c}+\frac{e^2\bm{A}^2(k\cdot x)}{2m_{\mathrm{e}}c},
\label{qrsfa1}
\end{equation}
and in the Volkov solution \cite{Ehlo,EJK},
\begin{align}
\psi^{(0)}_{\bm{p}}(x)=\frac{1}{\sqrt{V}}\exp\Bigl[-&\ii p\cdot x-\ii\int_0^{k\cdot x}\dd\phi'\Bigl(\frac{eA(\phi')\cdot p }{m_{\mathrm{e}}ck^0-\bm{p}\cdot\bm{k}} \nonumber \\
&-\frac{e^2A^2(\phi')}{2(m_{\mathrm{e}}ck^0-\bm{p}\cdot\bm{k})}\Bigr)\Bigr].
\label{qrsfa2}
\end{align}
Performing the same analysis as in the last two Sections, we can derive the expression for the differential probability distribution being analogous to Eqs.~\eqref{nsfa12} and~\eqref{rsfa16}. 
The relevant formulas are:
\begin{equation}
\bar{p}=p+\Bigl[-\frac{m_{\mathrm{e}}c\mu \langle f\rangle \varepsilon\cdot p}{m_{\mathrm{e}}ck^0-\bm{p}\cdot\bm{k}}+\frac{(m_{\mathrm{e}}c\mu)^2\langle f^2\rangle}{2(m_{\mathrm{e}}ck^0-\bm{p}\cdot\bm{k})}\Bigr]k
\label{nsfa4b}
\end{equation}
and
\begin{align}
G_{\mathrm{QRSFA}}(\phi)=\int_0^{\phi}\dd\phi' & \Bigl[\frac{m_{\mathrm{e}}c\mu \bigl(f(\phi)-\langle f\rangle\bigr) \varepsilon\cdot p}{m_{\mathrm{e}}ck^0-\bm{p}\cdot\bm{k}}
\nonumber \\
&+\frac{(m_{\mathrm{e}}c\mu)^2\bigl(f^2(\phi)-\langle f^2\rangle\bigr)}{2(m_{\mathrm{e}}ck^0-\bm{p}\cdot\bm{k})}\Bigr],
\label{nsfa5b}
\end{align}
with the corresponding Fourier coefficients $G_{\mathrm{QRSFA},N}^{(j)}$. Moreover, we obtain
\begin{align}
\mathcal{D}_{\mathrm{QRSFA}}(\bm{p})=&\sum_{N=-\infty}^{\infty}\frac{\ee^{2\pi \ii (\bar{p}^0-q^0-Nk^0)/k^0}-1}{\ii (\bar{p}^0-q^0-Nk^0)/k^0}
\nonumber \\
&\times\Bigl[\frac{G_{\mathrm{QRSFA},N}^{(1)} \varepsilon\cdot p}{m_{\mathrm{e}}ck^0}+\frac{m_{\mathrm{e}}c\mu G_{\mathrm{QRSFA},N}^{(2)}}{2m_{\mathrm{e}}ck^0}\Bigr]
\nonumber \\
&\times \tilde{\psi}_0(\bm{p}+(q^0-p^0)\bm{n}),
\label{nsfa12a}
\end{align}
and the dimensionless probability distribution of ionization (in atomic units) becomes
\begin{equation}
\mathcal{P}_{\mathrm{QRSFA}}(\bm{p})=\alpha^2\frac{(m_{\mathrm{e}}c)^4\mu^2}{(2\pi)^3}|\bm{p}|\cdot|\mathcal{D}_{\mathrm{QRSFA}}(\bm{p})|^2.
\label{nsfa12b}
\end{equation}
We will omit the derivation of the above expressions but, instead, we will discuss the physical nature of the substitutions \eqref{qrsfa1} and \eqref{qrsfa2}.

The first type of substitutions consists in modifying the electromagnetic fields and the vector potential such that they depend on both time and space variables. For the purpose of the current investigations, we call these the {\it retardation corrections}. It appears that the only modification imposed by these corrections is to shift the argument of the Fourier transform of the bound-state wavefunction $\tilde{\psi}_0(\bm{p})$ in~\eqref{nsfa9} by the vector proportional to the unit vector $\bm{n}$ and independent of the length of photon momentum [cf. Eq.~\eqref{nsfa12a}],
\begin{equation}
\bm{p}\longrightarrow \bm{p}+(q^0-p^0)\bm{n}.
\label{qrsfa3}
\end{equation}
The second type of substitutions, that are present only in the electron final state and for the purpose of our further discussion we call them the \textit{recoil corrections}, consists in modifying the denominators in~\eqref{qrsfa2} such that (cf. Ref.~\cite{Nord})
\begin{equation}
m_{\mathrm{e}}ck^0\longrightarrow m_{\mathrm{e}}ck^0-\bm{p}\cdot\bm{k}\approx p\cdot k.
\label{qrsfa4}
\end{equation}
This is equivalent to introducing into the nonrelativistic Keldysh theory the effective momentum-dependent mass,
\begin{equation}
m_{\mathrm{e}}\longrightarrow m_{\mathrm{eff}}=m_{\mathrm{e}}-\bm{p}\cdot\bm{n}/c,
\label{qrsfa5}
\end{equation}
which accounts for the electron recoil during the absorption of laser photons. Our numerical investigations presented below show that this is indeed the dominant contribution to the radiation pressure effects 
and that, in principle, the retardation corrections can be neglected in the QRSFA formulation of ionization, as they marginally contribute to the overall effect. 

Here, we also note that the recoil corrections are not introduced into the interaction Hamiltonian~\eqref{qrsfa1} for two reasons: 
firstly, they would lead to very small modifications of differential probability distributions and, secondly, such corrections would not appear 
in the length gauge since both the interaction Hamiltonian $-e\bm{\mathcal{E}}(\phi)\cdot\bm{r}$ and the gauge factor $\exp(\ii e\bm{A}(\phi)\cdot\bm{r})$ do not depend on the electron rest mass.

\section{Azimuthal distribution}
\label{sec:angle1}
\begin{figure}
\includegraphics[width=7cm]{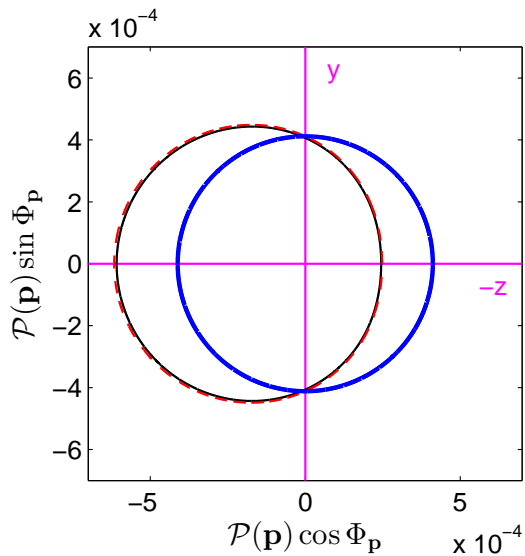}
\caption{(Color online) Azimuthal distribution of electrons ionized by the laser pulse defined in Sec.~\ref{sec:laser} with $N_{\mathrm{osc}}=4$ cycles and 
$\omega_{\mathrm{L}}=N_{\mathrm{osc}}\omega=1.55\,\mathrm{eV}$. The time-averaged intensity equals $I=2\times 10^{14}\,\mathrm{W/cm}^2$ and the polar angle of the electron ejection 
is $\Theta_{\bm{p}}=0.0475\pi$. The final electron energy is chosen to be $E_{\bm{p}}=m_{\mathrm{e}}c^2+30\,\mathrm{eV}$ for the RSFA and $\bm{p}^2/2m_{\mathrm{e}}=30\,\mathrm{eV}$ 
for the NSFA and QRSFA. The thick blue circle centered in the origin of the coordinate system corresponds to the nonrelativistic SFA, the thin black curve represents the results 
obtained within the RSFA, and the dashed thin red curve is for the QRSFA.
}
\label{pdn4i14Areiss1d20150625}
\end{figure}
\begin{figure}
\includegraphics[width=7cm]{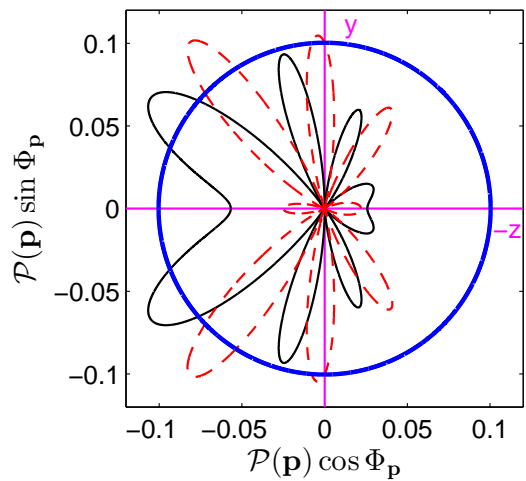}
\caption{(Color online) The same as in Fig.~\ref{pdn4i14Areiss1d20150625} but for $I=9\times 10^{15}\,\mathrm{W/cm}^2$, $\Theta_{\bm{p}}=0.02\pi$, and either $E_{\bm{p}}=m_{\mathrm{e}}c^2+200\,\mathrm{eV}$ 
or $\bm{p}^2/2m_{\mathrm{e}}=200\,\mathrm{eV}$.
}
\label{pdn4i14Hreiss1d20150625}
\end{figure}

As we have already mentioned, in the reference frame defined in Fig.~\ref{pedcone} the ionization probability distribution in the NSFA does not depend on the azimuthal angle $\Phi_{\bm{p}}$. 
This is illustrated in Fig.~\ref{pdn4i14Areiss1d20150625}, in which the NSFA is represented by the thick blue circle with the center located at the origin of the coordinate system. 
The results obtained in the RSFA and QRSFA are represented by the nearly identical closed curves (almost circles in their shapes) but with the center shifted towards the positive $z$-direction. 
This shift is the signature of the radiation pressure. We also observe that for the relativistic and quasi-relativistic SFA the results only marginally differ from each other. This means that, 
for the considered intensity of the laser field, it is not necessary to apply the full relativistic theory. In this case, the quasi-relativistic SFA very well describes the ionization distribution and 
effects related to the non-vanishing photon momentum. Therefore, one wonders at which laser intensities the QRSFA fails, showing significant deviations from the RSFA. To answer this question, 
in Fig.~\ref{pdn4i14Hreiss1d20150625} we compare all three approximations for much larger time-averaged intensity and a higher final kinetic energy of photoelectrons. As expected, 
the NSFA shows the rotational symmetry but both the QRSFA and RSFA exhibit much more complex structures, with many lobes that result from quantum interferences induced by the radiation pressure. 
It is also expected that the interference of probability amplitudes can enhance or suppress this effect, in the sense that the radiation pressure can either increase or decrease the  
photoelectron momentum along the propagation direction of the pulse. In the two figures presented in this Section, we have chosen the polar angles $\Theta_{\bm{p}}$ (for a given kinetic energy 
of photoelectrons) such that the averaged electron momentum parallel to the laser pulse propagation direction, i.e., to the $z$-direction, is positive. This agrees with our intuitive understanding 
of this phenomenon. However, with other choices of the polar angle $\Theta_{\bm{p}}$, the radiation pressure effect could be reversed and the probability distribution 
could be shifted towards negative $z$. Such possibilities will be discussed in Sec.~\ref{sec:angle3}.

A closer look at Figs.~\ref{pdn4i14Areiss1d20150625} and~\ref{pdn4i14Hreiss1d20150625} shows that on the $y$-axis the NSFA and QRSFA give almost the same results. If we analyze the corresponding 
expressions for the probability distributions we notice that for this particular direction, for which $\bm{n}\cdot\bm{p}=0$, they differ only by the momentum shift in the Fourier transform 
of the bound state wavefunction, i.e., by the retardation corrections. This suggests that the contributions to the radiation pressure effects originate mainly from the recoil corrections 
discussed in Sec.~\ref{sec:QRSFA}. In order to account for them properly it is not sufficient (or even not necessary) to replace $\bm{A}(\omega t)$ in the NSFA by $\bm{A}(\omega t-\bm{k}\cdot\bm{x})$, but 
modifications of the electron mass induced by the recoil corrections have to be considered. 

\begin{figure}
\includegraphics[width=7cm]{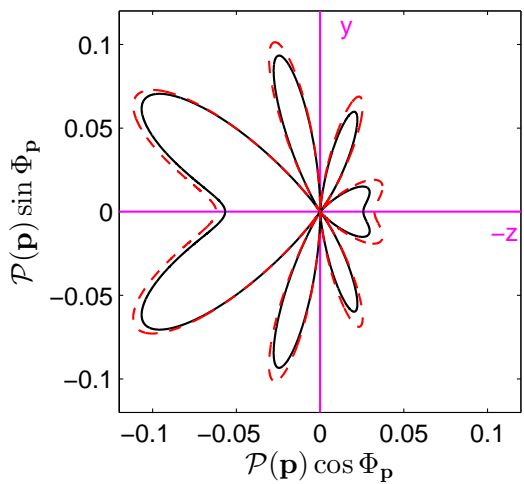}
\caption{(Color online) The same as in Fig.~\ref{pdn4i14Hreiss1d20150625} but omitting the NSFA results. In addition, the QRSFA (red dashed line) has been modified according to Eq.~\eqref{qrsfa6}. 
Higher recoil corrections, or even the full relativistic substitution $m_{\mathrm{e}}c\rightarrow p\cdot k/k^0$ in the final electron state, do not modify this figure.
}
\label{pdnv4i14Hreiss1d20150625}
\end{figure}

Further improvements of the QRSFA can be achieved by accounting for higher relativistic corrections to the effective mass in the electron final state, that go beyond the original Nordsieck substitution \eqref{qrsfa4},
\begin{equation}
m_{\mathrm{e}}c\longrightarrow m_{\mathrm{eff}}c=m_{\mathrm{e}}c-\bm{p}\cdot\bm{n}+\frac{\bm{p}^2}{2m_{\mathrm{e}}c}.
\label{qrsfa6}
\end{equation}
In Fig.~\ref{pdnv4i14Hreiss1d20150625} we compare the full relativistic theory with this version of the QRSFA. We observe a very good agreement between the RSFA and QRSFA results, which is by far better
than shown in Fig.~\ref{pdn4i14Hreiss1d20150625}. Note, that the retardation corrections contribute marginally to the QRSFA probability distribution and, 
for the considered in this paper laser pulse frequencies and intensities, they can be safely disregarded. For lower field intensities, the above extra modification of the QRSFA can be also neglected.

\section{Polar-azimuthal distribution}
\label{sec:angle2}

\begin{figure}
\includegraphics[width=7cm]{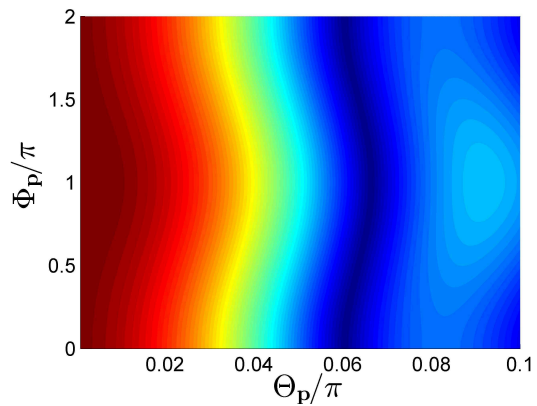}
\caption{(Color online) Color mapping of the polar-azimuthal distribution of ionization, treated in the RSFA, for the same parameters as in Fig.~\ref{pdn4i14Areiss1d20150625}. 
The QRSFA gives the same results whereas, for the NSFA, we observe the vertical straight stripes (i.e., the distribution is $\Phi_{\bm{p}}$-independent). The bending of these stripes 
is the signature of the radiation pressure.
}
\label{ptdA4i14p20150523r600}
\end{figure}

A more detailed insight into the radiation pressure effects can be gained by investigating the polar-azimuthal angle distributions of photoelectrons at a given final energy. When choosing this energy one has to remember about the domain of validity of the SFA, as discussed above. 
Note also that, for smaller laser intensities, one cannot select too large energies of 
final electrons since the corresponding probability distributions are very small and so they contribute marginally to the overall effect. For this reason, for the intensity 
$2\times 10^{14}\mathrm{W/cm}^2$ we choose the electron kinetic energy of $30\mathrm{eV}$, for which one could have doubts about the applicability of the Born approximation 
for the final scattering state of electrons. Nevertheless, it can provide an intuitive understanding of the radiation pressure effects. In Fig.~\ref{ptdA4i14p20150523r600} we present 
the color mapping of the polar-azimuthal distribution for the RSFA. Note, that for the QRSFA the map is almost identical, whereas for the NSFA the bended stripes become 
straight, as for the nonrelativistic case the corresponding probability distribution is $\Phi_{\bm{p}}$-independent. Because of the bended-type character of the relativistic distribution it is obvious now 
that, by choosing properly the polar angle of electron emission $\Theta_{\bm{p}}$, the laser pulse can exert on the photoelectron either the `positive' pressure, when 
the azimuthal probability distribution is shifted towards the positive $z$-axis (as it has been presented in Fig.~\ref{pdn4i14Areiss1d20150625}), or the `negative' pressure, when the azimuthal distribution 
is shifted in the opposite direction, for instance for $\Theta_{\bm{p}}=0.07\pi$. This problem is going to be quantified in the next Section.

\begin{figure}
\includegraphics[width=7cm]{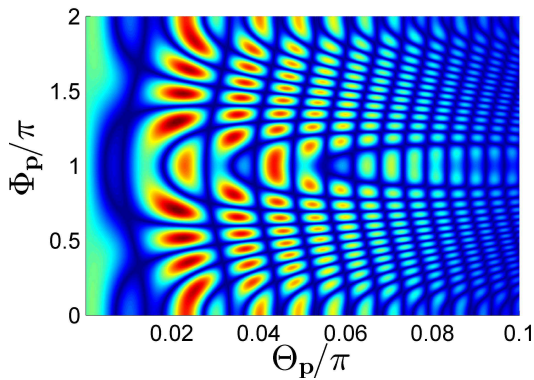}
\caption{(Color online) The same as in Fig.~\ref{ptdA4i14p20150523r600} but for the QRSFA and for parameters defined in the caption of Fig.~\ref{pdn4i14Hreiss1d20150625}. 
}
\label{ptd4i15Hn1p20150627r600}
\end{figure}
\begin{figure}
\includegraphics[width=7cm]{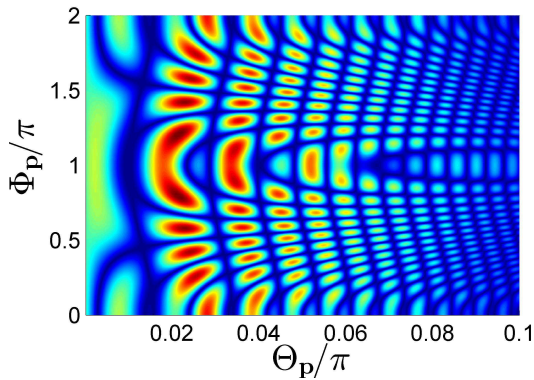}
\caption{(Color online) The same as in Fig.~\ref{ptd4i15Hn1p20150627r600} but for the RSFA.  
}
\label{ptd4i15Hp20150627r600}
\end{figure}

As for the azimuthal distribution discussed in the previous Section, for a larger pulse intensity and a higher electron kinetic energy, the polar-azimuthal distributions become much richer, 
as illustrated in Figs.~\ref{ptd4i15Hn1p20150627r600} and~\ref{ptd4i15Hp20150627r600}. This time we present the results for both the QRSFA and RSFA. Although, in details, these two distributions differ
from each other, they {\it do} show astonishing similarities. This indicates that, even for the time-averaged intensity equal to $9\times 10^{15}\mathrm{W/cm}^2$, the quasi-relativistic theory 
sufficiently well describes the radiation pressure effects. As before, the more intense laser pulses can exert on electrons both `positive' and 'negative' pressure 
depending on the selected value of the polar angle $\Theta_{\bm{p}}$.

\section{Azimuthal-angle-integrated distribution}
\label{sec:angle3}

The triply differential probability distribution $\mathcal{P}(E_{\bm{p}},\Theta_{\bm{p}},\Phi_{\bm{p}})$, which has been derived above, can be used to determine the averaged momentum of ejected electrons,
\begin{equation}
\langle \bm{p}\rangle=\int\frac{\dd E_{\bm{p}}}{\alpha^2m_{\mathrm{e}}c^2}\dd^2\Omega_{\bm{p}}\, \bm{p}\mathcal{P}(E_{\bm{p}},\Theta_{\bm{p}},\Phi_{\bm{p}}).
\label{azimuth1}
\end{equation}
For the chosen polarization and propagation direction of the driving pulse, this probability distribution fulfills the symmetry condition,
\begin{equation}
\mathcal{P}(E_{\bm{p}},\Theta_{\bm{p}},2\pi-\Phi_{\bm{p}})=\mathcal{P}(E_{\bm{p}},\Theta_{\bm{p}},\Phi_{\bm{p}}).
\label{azimuth1a}
\end{equation}
This leads to the conclusion that the $y$-component of the averaged momentum has to vanish, $\langle p_y\rangle=0$. Since we consider a finite pulse, the shape function of the electromagnetic potential $f(\phi)$ 
has, in general, a non-vanishing constant term in its Fourier decomposition. This means that the $x$-component of the averaged momentum $\langle p_x\rangle$ may not vanish. 
This effect cannot be attributed to the radiation pressure, as it also appears in the NSFA. This is in contrast to the non-vanishing $z$-component of the averaged
momentum $\langle p_z\rangle$. The latter, in the light-cone coordinates denoted as $\langle p_z\rangle=\langle p^{\|}\rangle$, is equal to
\begin{equation}
\langle p^{\|}\rangle=\int\frac{\dd E_{\bm{p}}}{\alpha^2m_{\mathrm{e}}c^2}\dd^2\Omega_{\bm{p}}\,  \mathcal{P}(E_{\bm{p}},\Theta_{\bm{p}},\Phi_{\bm{p}})(-|\bm{p}|\sin\Theta_{\bm{p}}\cos\Phi_{\bm{p}}).
\label{azimuth2}
\end{equation}
It can be represented as the integral
\begin{equation}
\langle p^{\|}\rangle=\int\frac{\dd E_{\bm{p}}}{\alpha^2m_{\mathrm{e}}c^2}\dd \Theta_{\bm{p}}\,  |\bm{p}|W^{\|}(E_{\bm{p}},\Theta_{\bm{p}}),
\label{azimuth3}
\end{equation}
with the two-dimensional distribution,
\begin{equation}
W^{\|}(E_{\bm{p}},\Theta_{\bm{p}})=-\sin^2\Theta_{\bm{p}}\int_0^{2\pi}\dd\Phi_{\bm{p}}\, \cos\Phi_{\bm{p}}\mathcal{P}(E_{\bm{p}},\Theta_{\bm{p}},\Phi_{\bm{p}}).
\label{azimuth4}
\end{equation}
The advantage of introducing this type of distribution is that it is equal to 0 for the NSFA, when no radiation pressure effects are expected. In addition, it quantifies the notions of the 
`positive' and `negative' radiation pressure discussed above, as they are strictly related to the sign of $W^{\|}(E_{\bm{p}},\Theta_{\bm{p}})$.

\begin{figure}
\includegraphics[width=7cm]{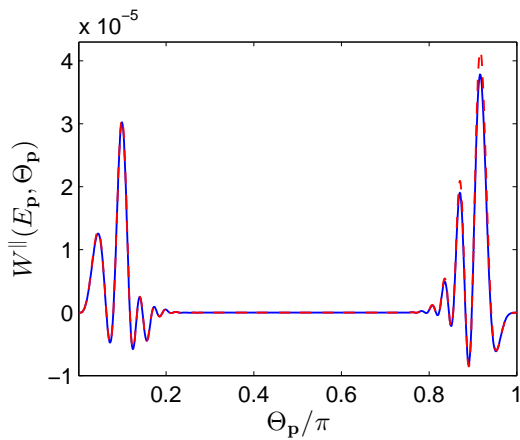}
\caption{(Color online) The probability distribution~\eqref{azimuth4} of measuring the electron momentum along the pulse propagation direction. 
The parameters are the same as in Fig.~\ref{ptdA4i14p20150523r600}. The solid blue line corresponds to the RSFA and the dashed red line to the QRSFA.
}
\label{ptd2x14d20150729tot2}
\end{figure}
\begin{figure}
\includegraphics[width=7cm]{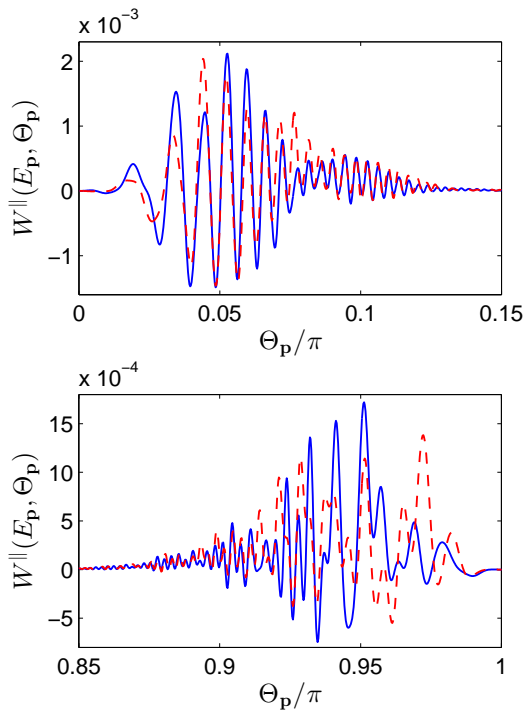}
\caption{(Color online) The same as in Fig.~\ref{ptd2x14d20150729tot2} but for parameters defined in the caption of Fig.~\ref{pdn4i14Hreiss1d20150625}. 
For larger intensities we observe discrepancies between the RSFA and QRSFA, as compared to the results presented in Fig.~\ref{ptd2x14d20150729tot2}.
Note that with the extra substitution~\eqref{qrsfa6} these discrepancies become marginally small.
}
\label{ptd9x15d20150729tot2}
\end{figure}

In Fig.~\ref{ptd2x14d20150729tot2}, we present $W^{\|}(E_{\bm{p}},\Theta_{\bm{p}})$ for $E_{\bm{p}}=m_{\mathrm{e}}c^2+30\mathrm{eV}$ (for RSFA), $\bm{p}^2/2m_{\mathrm{e}}=30\mathrm{eV}$ 
(for QRSFA), and for the pulse parameters considered in Fig.~\ref{ptdA4i14p20150523r600}. As expected, these distributions are almost identical. Although, predominantly, they are positive 
(i.e., the laser pulse exerts the `positive' pressure on electrons), they also take negative values. If, however, we integrate $W^{\|}(E_{\bm{p}},\Theta_{\bm{p}})$ 
over the polar angle $\Theta_{\bm{p}}$ we obtain the positive value. This means that electrons of this particular energy are pushed in the positive $z$-direction by the pulse. 
A similar pattern is observed for larger pulse intensities and higher electron energies (see, e.g., Fig.~\ref{ptd9x15d20150729tot2}), but now there are more oscillations of 
$W^{\|}(E_{\bm{p}},\Theta_{\bm{p}})$ and the discrepancies between the RSFA and QRSFA results are clearly visible. Despite the obvious differences, qualitatively both distributions look similar. 
Note that, even though for different energies and polar angles the laser pulse can affect the electron distribution differently (i.e., it can exert either 'positive' or 'negative' pressure
on the electron), its action on the whole system 'nucleus+electron' is always 'positive'.

\section{Conclusions}
\label{sec:Conclusions}

We have studied the effects related to the radiation pressure exerted by short, linearly-polarized laser pulses on hydrogen atoms during their ionization. 
In doing so, we have presented three types of the SFA developed for the velocity gauge. We have shown that, for the Ti-Sapphire laser pulses of intensities up to $10^{16}\mathrm{W/cm}^2$, 
the quasi-relativistic SFA predicts nearly identical effects as the full relativistic SFA for the Dirac equation. In particular, we have discussed modifications related to the retardation 
and recoil corrections. We have shown that the dominant contribution to the radiation pressure originates from the recoil corrections to the nonrelativistic SFA. These findings create
a possibility to modify the complex-time trajectory method for ionization such that it can account for the radiation pressure effects. Also, we have demonstrated that, depending on 
photoelectrons final energies and polar angles, the radiation pressure effects can be of `positive' or `negative' types. They are more pronounced in the energy-angular distributions 
of ejected electrons than in the integrated, parallel (with respect to the laser pulse propagation direction) momentum distributions, where the overall effect is significantly suppressed. 
Extensions of the theoretical methods considered here to the length gauge and arbitrary laser pulse polarizations are under development and are going to be presented in due course.

To summarize, we have demonstrated that, for short laser pulses of time-averaged intensities not larger than $10^{16}\mathrm{W/cm}^2$ with the carrier frequency of $1.55\mathrm{eV}$, 
the radiation pressure effects in ionization can be well described by the nonrelativistic Keldysh theory provided that the electron mass is replaced by the effective momentum-dependent one. 
The latter accounts for the recoil of electrons during the absorption of laser photons. This has been shown by analyzing differential probability distributions in the nonrelativistic and relativistic 
strong-field approximations in the velocity gauge, but we expect that the same conclusion is also valid for the length gauge.

\section*{Acknowledgements}

This work is supported by the Polish National Science Center (NCN) under Grant No. 2012/05/B/ST2/02547. 
K.K. acknowledges also the support from the Kosciuszko Foundation.


\begin{thebibliography}{99}

\bibitem{press0}
P. Lebedew, Ann. d. Phys. {\bf 311}, 433 (1901).

\bibitem{press1}
C. T. L. Smeenk, L. Arissian, B. Zhou, A. Mysyrowicz, D. M. Villeneuve, A. Staudte, and P. B. Corkum, Phys. Rev. Lett. {\bf 106}, 193002 (2011).

\bibitem{press2}
A. S. Titi and G. W. F. Drake, Phys. Rev. A {\bf 85}, 041404(R) (2012).

\bibitem{press3}
H. R. Reiss, Phys. Rev. A {\bf 87}, 033421 (2013).

\bibitem{Chelkowski}
S. Chelkowski, A. D. Bandrauk, and P. B. Corkum, Phys. Rev. Lett. {\bf 113}, 263005 (2014).

\bibitem{press4}
I. A. Ivanov, Phys. Rev. A {\bf 91}, 043410 (2015).

\bibitem{ritus1}
V. I. Ritus, Trudy FIAN {\bf 111}, 5 (1979).

\bibitem{ritus2}
V. I. Ritus, J. Sov. Laser Res. {\bf 6}, 497 (1985).

\bibitem{FKK} 
F. Ehlotzky, K. Krajewska, and J. Z. Kami\'nski, Rep. Prog. Phys. \textbf{72}, 046401 (2009).

\bibitem{PiazzaRev} 
A. Di Piazza, C. M\"uller, K. Z. Hatsagortsyan, and C. H. Keitel, Rev. Mod. Phys. {\bf 84}, 1177 (2012).

\bibitem{Keldysh}
L. V. Keldysh, Zh. Exp. Theor. Fiz. {\bf 47}, 1945 (1964).

\bibitem{Perelomov}
A. M. Perelomov, V. S. Popov, and M. V. Terent'ev, Zh. Exp. Theor. Fiz. {\bf 50}, 1393 (1966).

\bibitem{Gribakin}
G. F. Gribakin and M. Yu. Kuchiev, Phys. Rev. A {\bf 55}, 3760 (1997).

\bibitem{Karnakov}
B. M. Karnakov, V. D. Mur, S. V. Popruzhenko, and V. S. Popov, Phys. Usp. {\bf 58}, 3 (2015).

\bibitem{Popruzh}
S. V. Popruzhenko, J. Phys. B {\bf 47}, 204001 (2014).

\bibitem{CKK}
F. Cajiao V\'elez, K. Krajewska, and J. Z. Kami\'nski, Phys. Rev. A {\bf 91}, 053417 (2015).

\bibitem{Faisal}
F. H. M. Faisal, J. Phys. B {\bf 6}, L89 (1973).

\bibitem{Reiss}
H. R. Reiss, Phys. Rev. A {\bf 22}, 1786 (1980).

\bibitem{Sujata1}
F. H. H. Faisal and S. Bhattacharyya, Phys. Rev. Lett. {\bf 93}, 053002 (2004).

\bibitem{eikHeidelberg}
M. Klaiber, E. Yakaboylu, and K. Z. Hatsagortsyan, Phys. Rev. A {\bf 87}, 023418 (2013).

\bibitem{eikHeidelberg2}
E. Yakaboylu, M. Klaiber, and K. Z. Hatsagortsyan, Phys. Rev. A {\bf 91}, 063407 (2015).

\bibitem{Kam}
J. Z. Kami\'nski, Acta Phys. Pol. A {\bf 66}, 517 (1984).

\bibitem{Volkov}
D. M. Wolkow, Z. Phys. {\bf 94}, 250 (1935).

\bibitem{KKC}
K. Krajewska and J. Z. Kami\'nski, Phys. Rev. A {\bf 85}, 062102 (2012).

\bibitem{KKBW}
K. Krajewska and J. Z. Kami\'nski, Phys. Rev. A {\bf 86}, 052104 (2012).

\bibitem{KK1}
K. Krajewska and J. Z. Kami\'nski, Laser Phys. Lett. {\bf 11}, 035301 (2014).

\bibitem{BjorkenDrell}
J. D. Bjorken and S. D. Drell, \textit{Relativistic Qauntum Mechanics} (McGraw-Hill, New York, 1964).

\bibitem{Ehlo}
F. Ehlotzky, Can. J. Phys. {\bf 63}, 907 (1985).

\bibitem{EJK}
F. Ehlotzky, A. Jaro\'n, and J. Z. Kami\'nski, Phys. Rep. {\bf 297}, 63 (1998).

\bibitem{Nord}
A. Nordsieck, Phys. Rev. A {\bf 93}, 785 (1954).



\end{thebibliography}
\end{document}